# Legal routes for accomplishing corporate environmental compliance against the "carbon peaking and carbon neutrality" goals

Yedong Zhang, Hua Han

**Abstract:** Against the macro-background of "carbon peaking and carbon neutrality" goals, eco-environment protection regulations are increasingly stricter. Facing high government regulatory risks and frequent environment lawsuits, corporate environmental compliance starts to play a vital role in healthy corporate operation. Law fulfillment routes constitute a critical part in corporate environmental compliance. Few academic scholars have conducted a profound analysis or discussion of legal accomplishment routes for corporate environmental compliances. As a matter of fact, legal routes for accomplishing corporate environmental compliance should be based proper theories concerning corporate environmental rights and obligations as well as dual layer nested governance structure (government environmental power and corporate environmental liabilities). Under the guidance of environmental jurisprudence, enterprises are responsible for setting up practical legal fulfillment routes for their environmental compliance-related rights and obligations. A diversified environmental governance layout composed of government regulation, enterprise self-discipline and social participation should be established. Within enterprises, effective legal routes should be developed for dealing with government regulatory risks and environment lawsuit risks at the same time.

## Introduction

In March 2021, in the 9th meeting of the Central Finance and Economics Committee, it was proposed to include carbon peaking and carbon neutrality into "the overall layout of ecological civilization construction". Therefore, it becomes an inevitable requirement to reinforce corporate environmental compliance mechanism construction in accomplishing "carbon peaking and carbon neutrality" goals. Corporate environmental compliance starts to play a critical role in standard and healthy corporate operation. The demand for corporate environmental compliance is unprecedently eager now.[1] Against the macro-background of carbon peaking and carbon neutrality goals, our country has gradually shifted from "dual control on total energy consumed

---

Corresponding author: Yedong Zhang, Associate Professor at Law School of Shenzhen University and Research Associate at the Shenzhen University Institute of Science, Technology, and Law.
Email addresses: 21110270022@m.fudan.edu.cn (Z. Yedong), hhuaview@163.com (H. Han).

[1] David Waskow, Chris Jochnick & Juge Gregg, Progressive Approaches to International Environmental Compliance, 15 FORDHAM ENVTL. L. REV. 496 (2004).

and energy consumption intensity" to "dual control on total carbon emission and intensity". A rising tendency has been witnessed in the prices of related commodities. Enterprises are facing much higher environmental regulatory pressure. A number of them are forced to reduce production or even be closed due to non-compliance with relevant environmental laws and regulations. This severely undermines corporate production and operation. As few corporates could avoid causing environmental pollution and ecological damage during their development, corporate environmental compliance is increasingly important in helping corporates in dealing with possible government regulation and environment lawsuits.[2] Previous works and papers concerning corporate compliance either focus too much on discussing corporate environmental compliance alone to conduct a categorized collation or analyze the corporate environmental compliance from management perspective instead of legal perspective without giving any compliance suggestions and strategies. It's usually too late when enterprises find themselves caught in environmental non-compliance.[3] In view of this, it is urgent now to build legal routes for corporate environmental compliance in the future. [4]Next the authors are going to analyze how to build legal accomplishment routes of corporate environmental compliance in three parts. In the first place, existing environmental risks faced by enterprises are categorized to demonstrate prerequisites for corporate environmental compliance. Then, an analysis is performed on jurisprudence structure of corporate environmental compliance to determine the fundamental jurisprudence for enterprises to conduct environmental compliance work. At last, based on the aforesaid work, we set out to build legal accomplish routes for domestic corporate environmental compliance in the government-enterprise-society common governance structure.

## I. Prerequisites for corporate environmental compliance: environmental risk response

Corporate environmental compliance risk is the risk of legal liability, regulatory punishment, financial and reputation loss or even bankruptcy faced by an enterprise after violating environmental protection laws or regulations.[5] The emerging multi-governing model in environmental protection could be viewed as a kind of benign interaction among three parties, namely duteous government, law-abiding enterprise and active public.[6] However, practice reveals those three parties appear insufficient in fulfilling their governance duties so that they can hardly complete the environmental governance tasks. Apart from that, government regulation force lacks standardization, and disorderly public participation becomes prominent. [6]All those factors bring in dual environmental risks for enterprises in their production and management, namely government

---

[2] Richard F. Charfield-Taylor, Environmental Compliance: Negotiating the Regulatory Maze, 3 MO. ENVTL. L. & POL'y REV. 3 (1995).
[3] Wanyi Zhao, Peng Wang, 'On the Legal Realization Route of Synthetical and Harmonized Regulating Corporates Compliance Behaviors in China'(2021) 7 Hebei Law Science 39(in Chinese).
[4] Ruihua Chen, Basic Theory of Enterprise Compliance(2nd edn, Law Press 2021)103-65(in Chinese).
[5] Timothy Riley and Cai Huiyan, Unmasking Chinese Business Enterprises: Using Information Disclosure Laws To Enhance Public Participation In Corporate Environment Decision Making, 33 Harv. Envtl. L. Rev. 177.
[6] Yang Chen, 'On the logic of check and balance in China's environmental multi-governance',(2020) 6 Chinese Journal of Population, Resources and Environment 30(in Chinese).

regulation and environment lawsuit.

## (I) Government regulation risk: realistic basis and practice logic

In recent years, with proposal of "carbon peaking and carbon neutrality" goals, corporate environmental compliance is shifting from "dual control on total energy consumed and energy consumption intensity" to "dual control on total carbon emission and intensity". The practical and frequent legal risks including administrative punishment, production suspension and even criminal accountability faced by enterprises cannot be ignored now. As environmental protection turmoil becomes normalized, it turns out to be especially important and urgent now to enhance the preventive measures against government regulation risk. "Immense tension and profound crack".[7] Corporate environmental compliance is both a kind of protection of the development right enjoyed by enterprises and an urge on enterprises to perform their environmental protection obligations. Government regulation on enterprises serves as an important propellant for the latter to carry out environmental compliance work. Nevertheless, this also has its own complicated objective background, including realistic basis of government's over-regulation and practice logic of regulation violation by enterprises.

### 1. Realistic basis of over-regulation on behalf of the government

The realistic basis of government's over-regulation lies in the sweeping or radical approach in environmental governance. Over-regulation by government may leave the impression that the government "fails to do well in environmental regulation regardless of its capabilities". To be more specific, under constant intense pressure from the state in environmental protection, local governments have to choose "sweeping" approach. The "sweeping" over-regulation by the government represents "radicalism" in environmental governance. We may take the compulsory power switch-out by many local governments as an example. In many places, widespread power switch-out is applied, which reveals incompetency in environmental and economic governance. Due to an ambivalent attitude in synergizing local environmental governance with economic development, some local governments have failed to assign carbon emission quotas to economy, livelihood and environment in a reasonable way in their long-term development plan. Consequently, their daily governance is still in the stage of "symptomatic treatment". When some regions are after short-term economic effect only so that they acquiesce full-capacity production by some enterprises and industries and local carbon emission are beyond permissible limits for long time. In such case, they have to take some measures such as compulsory power switch-out to meet "carbon peaking and carbon neutrality" goals prescribed by the central government. This belongs to typical symptomatic governance that grasps no essence of an issue. It could only aggravate local carbon emission and reduce regional economic structure to be a deformed development featuring low-level, high-energy consumption and heavy pollution. It also brings damage to local environment and public livelihood.

---

[7] Bowen, H. R . Social Responsibilities of the Business Man [J]. Harper Brothers, 1953(6).

**2. Practice logic of supervision violation by enterprises**

The practice logic underlying supervision violation by enterprises is their inborn pursuit of benefits and expansion. To maximize profit and "be systematically irresponsible" for public welfare is the exact cause for enterprises in polluting the environment and damaging the ecology.[8] As "rational economic-man", an enterprise is destined to maximize its benefit. This is simply unchangeable. Traditional enterprises' only goal is to maximize the profit. The unicity in corporate goal makes them turn a blind eye to environmental liabilities. [9]Driven by profit-pursuing nature, some enterprises even take illegal measures such as stealthy emission or discharge to evade their due environmental protection obligation and pollution treatment liability. As a co-governor in environmental issues, some third-party service suppliers may commit illicit behaviors that are against public interest in environment due to their profit-seeking nature. Now the widespread environmental illegality cases in different provinces and cities highlight the stunning number and scope of environmental non-compliance issues caused by enterprises. The high number of environmental administrative punishment cases reflects the fact that profit-seeking nature of enterprises is still a significant reason hindering performing of environmental governance obligation. Regardless of enforcement of a series of "record strict" environmental protection laws and regulations such as amended Environmental Protection Law and Air Pollution Prevention and Control Act since 2015, enterprises that "challenge the law" are never rare.

**(II) Environmental lawsuit risk: "Tyranny of the majority" and "public opinion manipulation"**

At present, social organizations are still rather weak in power. On the other hand, the non-disclosure and non-transparency in corporate environmental information leads to disorderly participation of social public in environmental governance. As social public changes from supervisor to co-governor, its discourse power and influence in this regard would also be enhanced. However, due to heterogeneity of interest in the composition of social public, intergroup benefit conflict and multiple appeals could easily cause social public participation to become disturbance and such further social participation crises as "tyranny of the majority" and "public opinion manipulation".[10] In such situations, the original public interest gets translated to be satisfaction of personal interest. Current legal protection against environmental tort is gradually transformed from private environmental interest protection to dual protection of both public and private environmental interest. It suggests the environmental lawsuit risks faced by enterprises could be divided into public environmental lawsuit and private environmental lawsuit.

---

[8] Binglin Tan, 'On the Government's Management-Based Regulation of Enterprises'(2019) 6 Journal of Jurists(in Chinese).

[9] Donald A. Carr, William L. Thomas, Devising a Compliance Strategy Under the ISO 14000 International Environmental Management Standards, 15 Pace Envtl. L. Rev. 85.

[10] Qian Zhu, Jingjing Yu, 'Research on the approach to the Connection of Environmental Civil Public and Private Interest Litigation from the Perspective of Civil Code'(2021) 5 Journal of China University of Geosciences (social sciences edition) 21(in Chinese).

### 1. Risk of public-interest environmental lawsuit

As one of the democratic game scenarios, "tyranny of the majority" could also be traced in the public participation in environmental issues. The term "public" is in itself a complicated composition filled with interest distinguishment and conflicts. When it comes to environmental protection, the "public" can be divided into two categories, namely the "uninterested public" and the "interested public". Social environmental protection organizations are in the former category, representing social public to take part in environmental governance and supervise environmental protection by enterprises.[11] Nevertheless, in reality, scarce environmental protection organizations could satisfy the lawsuit prerequisites in Article 55 of Civil Procedure Law and Article 58 of Environmental Protection Act. Appearing quite weak, those environmental protection organizations could hardly form orderly, organized and procedure-legitimate supervision mechanisms. With social organizations being absent in supervision, enterprises are not effectively restricted when violating environmental protection regulations so that they are not warned promptly by related environmental protection organizations about controlling environmental polluting behaviors in spite of the already extremely severe environmental pollution and ecological damage they have committed. In consideration of the dilemma of filing public-interest lawsuit by environmental protection organizations, the procuratorate starts undertaking the function of filing such a lawsuit. In this case, the procuratorate may be viewed as an uninterested special organization. In accordance with Article 55 of Civil Procedure Law, if no environmental protection organization files the public interest lawsuit, the lawsuit may be initialized by a procuratorate. The public interest lawsuit concerning environmental protection initialized by a procuratorate could be deemed as a supervision mechanism in some sense. It represents a sort of public power supervision on corporate environmental non-compliance. In actual practice and operation, another problem arises. When public interest lawsuit is filed by the procuratorate, social environmental protection organizations become inevitably increasingly weak. Without the buffering effect of social environmental protection organizations, as long as a procuratorate takes measures against environmental non-compliance, the enterprise involved must have already committed very serious acts. An environmental crime may be constituted already. In view of the analysis above, public interest lawsuit filed by uninterested public poses a significant threat to the survival and development of enterprises. Active environmental compliance measures should be immediately taken by the enterprises involved.

### 2. Risk of private-interest environmental lawsuit

Private environmental interest lawsuit against an enterprise is filed directly by interested social public. When facing grievous interest damage, hysterical social public are susceptible to initializing mass disturbance and thus causing "public opinion manipulation" to the enterprise. In such mass disturbance over environmental issues, the government and enterprise involved may have to make considerable compromise and concession under public opinion pressure. However, afterwards a calm and objective review of some construction projects may easily bring in such findings that excessive exaggeration does exist in some cases. But the construction projects are already cancelled, which leads to a series of negative effects. In such case, the government, the enterprise involved and the public "all lose". There is no winner at all. Public opinion manipulation is manifested in private environmental interest lawsuit. Even if there is only one subject of litigation, the resulting "public opinion manipulation" would turn to be really frightful if let be magnified by the public opinion without constraints. In accordance with Article 1229 of the Civil Code, the infringed shall be entitled to file a private-interest environmental lawsuit. In such a situation, the environmental infringer shall undertake the no-fault liability.[12] Since strict accountability principle applies to the enterprise involved in a private-interest environmental lawsuit, the infringed public is inclined to attribute all their environmental interest damages to the original sin of the enterprise and the connivance by the government so that they may take aggressive act which not only is unhelpful for problem solving but also may further complicate the problem. The enterprise reputation would be greatly undermined so that normal production and operation is affected. Therefore, private-interest environmental lawsuit constitutes another material risk for enterprises, because it could severely undermine the enterprises' survival and development. Enterprises must take corresponding environmental compliance strategies to avoid being caught in such situations. Since environmental risks looming around the enterprises may bring in economic loss, reputational damage or direct threat to enterprise survival and development, it is really urgent for enterprises to conduct environmental compliance work. Few scholars have set out to demonstrate the jurisprudential basis for corporate environmental compliance. Thus, the authors are going to have a tentative analysis of the jurisprudential structure of corporate environmental compliance.

---

[11] Yixiang Xu, 'The binary structure of public participation rights: Focusing on the legal norms of public participation in the environmental administration'(2018) 2 Journal of Central South University (Social Sciences Edition) 24(in Chinese).

## II. Basis for corporate environmental compliance: a novel jurisprudential structure

To demonstrate the basis for corporate environmental compliance, it is inevitable to make a jurisprudential analysis. Discussion of right and obligation is within the legal relation category between private entities, namely private jurisprudence, while that of authority and responsibility is within the legal relation category between public power body and private entity. The difference between environmental laws and traditional law lies in the fact that the involved interest is a mix of public and private interests, which could be summarized as public-and-private interest. When looked from the right-obligation perspective, enterprises enjoy environmental rights and shoulder environmental obligations. And an analysis from power-responsibility perspective reveals the government is entitled to supervise enterprises whereas enterprises must undertake environmental liabilities. As for specific forms, the responsibilities include civil responsibilities, administrative responsibilities and even criminal responsibilities. Corporate environmental compliance should not be confined to traditional distinction between public and private jurisprudence. Instead, a novel environmental jurisprudential structure should be set up for corporate's environmental compliance.

---

[12] Qian Zhu, Jingjing Yu, 'Research on the approach to the Connection of Environmental Civil Public and Private Interest Litigation from the Perspective of Civil Code'(2021) 5 Journal of China University of Geosciences (social sciences edition) 21(in Chinese).

### (I) Private jurisprudential structure of corporate environmental compliance: right-obligation structure

**1. Corporate environmental right**

Corporate environmental right means the right legally endowed on such organizations as enterprises and units to enjoy pleasant environment and reasonably use environmental resources. Right is the cornerstone of modern law philosophy, the theoretical expression of which is "right-based theory".[13] Thus, we may say environmental right is the cornerstone of environment law. Environmental right incorporates the right of the state, legal person and citizen to use and enjoy natural environmental conditions.[14] As specified in Article 57 and 76 in the Civil Code, a corporate is a profit-making corporation. It is of no doubt that corporates are entitled to enjoy corporate environment. We can also derive from the aforesaid facts that the right enjoyed by corporates in accessing pleasant environment and reasonably using environmental resources is the environmental right between citizen's environmental right and state's environmental right. This right plays a special role in bridging those two rights. In previous studies, some scholars deny the environmental right for corporates,[15] which cannot be shared by the authors for following reasons. First of all, as a legal person, corporate can of course be an entity of environmental right. Secondly, it is an important task for the Civil Code drafted in the new era by our country to guarantee a friendly and pleasant living environment and instruct reasonable and efficient utilization of natural resources. Corporate is undoubtedly a civil entity that is endowed with environmental right in civil sense just like other civil rights such as right of name, right of reputation and right of honor as prescribed in the General Provisions of Civil Law. Thirdly, to the knowledge of the authors, the statement denying corporate environmental right contains two flaws: it violates the basic rule of the unity of right and obligation and it cannot explain the trading nature of environmental right and interest. Last but not the least, others declare corporate environmental right theory is purely Western legal theory that cannot adapt well to Chinese conditions and a more proper practice is to substitute corporate environmental right theory with corporate environmental obligation theory in public law. But it fails to take into account the social environment change factor. As an entity of environmental right, corporates are entitled to have an access to environmental right. The objective of offering corporates an access to those environmental resource rights is to guarantee a sustainable development of the whole society.

---

[13] Wenxian Zhang, Research on the Category of Legal Philosophy(1st edn, China University of Political Science and Law Press 2001 )399-281(in Chinese).

[14] Rena I. Steinzor, REINVENTING ENVIRONMENTAL REGULATION: THE DANGEROUS JOURNEY FROM COMMAND TO SELF-CONTROL, 22 Harv. Envtl. L. Rev. 103.

[15] Somendu B. Majumdar, Voluntary Environmental Compliance Auditing: A Primer, 7 Fordham Envtl. Law J. 817.

After having decided the lawfulness of corporate environmental right, we're going to talk about the composition of corporate environment right. In the current Environmental Protection Law, Article 22 and 23 specify that enterprises could enjoy certain environmental interests due to their contribution to environmental protection.[16] When examined from right-interest perspective, this could be an embodiment of corporate environmental right. On the other hand, corporate environmental right is a bit different from traditional civil rights, because corporates have both the right to enjoy pleasant environment and the obligation to protect it. Should environmental right be confirmed as a fundamental right in our Constitution and a civil right accessible to the natural person, legal person and unincorporated organization in the civil law system, a top-down environment right system from the Constitution of at the highest order should be established. This would be of great realistic significance for environmental right to shift from obligatory right to actual one. Corporate is of course a civil entity that has an access to environmental right in civil sense just like they are entitled to enjoy other civil rights such as the right of name, right of reputation and right of reputation as prescribed in the General Provisions of Civil Law.

### 2. Corporate environmental obligation

Corporate environmental obligation is a general obligation to care for and protect environmental resources on behalf of such organizations as enterprises and units. [17]It is inclusive of the obligations to protect and improve environment, prevent environmental pollution and eliminate other public nuisances. Details about the corporate environmental obligation are scattered around in the current Environmental Protection Law. In this Law, it is specified in Article 6 that "all the units and individuals are obliged to protect the environment." As listed in following table, statutory corporate obligation covers four aspects, namely clean production, emission reduction and lawful pollution discharge, environmental management and obedience to relevant regulation and supervision. For clean production, clean energy as well as process and equipment with higher resource utilization rate and less pollutant discharge should be preferred. To reduce emission and legally discharge pollutants, illegal pollution discharge evading regulation by use of concealed passage should be avoided to prevent possible pollution and hazard. Emission and discharge should comply with sewage discharge permit without surpassing quotas and total volumes. As for environment management, proper detection devices should be installed and put into use and environmental emergency plan should be drafted. Obedience to regulation and supervision includes approval before construction, onsite inspection and disclosure of pollution discharge information details.

[16] Donald A. Carr and William L. Thomas, Devising a Compliance Strategy Under the ISO 14000 International Environmental Management Standards, 15 Pace Envtl. L. Rev. 85.
[17] Shouqiu Cai,'Exploration of Environmental Rights'(1982) 3 Chinese Social Sciences(in Chinese).

（1）Clean production obligation

Corporates' clean production obligation means corporate should prefer use of clean energy as well as process and equipment with higher resource utilization rate and less pollutant discharge. Integrated waste utilization technology and pollutant detoxification technology should be adopted to cut down pollutant generation. Corporates constitute the basic organizations and drive force of clean production implementation. Article 40 of Environmental Protection Law stipulates the enterprises' right and obligation in clean production in order to give a play to their proactivity and creativity in this aspect. While being obliged to submit reports and data concerning clean production, enterprises are also entitled to gain clean production information, materials, capital and technical assistance from the government.

（2）Obligation of emission reduction and lawful pollution discharge

This obligation may be divided into two layers, namely emission reduction and lawful pollution discharge. The first layer is about cutting down pollutant discharge and carbon dioxide emission, preventing and alleviating environmental pollution and ecological damage, and bearing the liability for polluting the environment or bringing damage to the ecology. For the second layer, it means enterprises should obey pollutant treatment regulations, obtain pollutant discharge permit in a lawful and legal way, and pay sewage discharge fee or environmental protection tax as obliged.

（3）Environmental management obligation

Corporate environmental management obligation means a corporate should set up environmental protection liability system to define responsibilities of both the head and related staff in its production and operation. This obligation is made up of environment supervision and pollution contingency plan. To supervise the environment, enterprises with significant pollutant discharge should establish corresponding system and use adequate supervision equipment. They are required to strengthen environment monitoring report and institutionalize environmental monitoring data and information management to make sure about an efficient communication of environmental monitoring information. This is helpful in improving promptness, specificity, accuracy and systematicity of environmental decision-making and management service. Under pollution contingency report obligation, corporates should draft environmental contingency and take measures and make report promptly in case of meeting such an accident. To be more specific, when an accident or emergency occurs, in which environment may be or is already severely contaminated and people's lives and properties are at great risk, the enterprise involved is legally obliged to make report about details and take corresponding countermeasures. It is clearly defined in our Environmental Protection Law that the unit or individual that may cause or already caused environmental pollution due to any accident or emergency should immediately report the case to local administrative agency in charge of environmental protection and other related bodies and actively cooperate with subsequent accident investigation. Here the aforesaid "pollution accident" refers to any emergency that causes significant pollution to the environment, brings harm to people's life and health, material loss to social economy and residents' properties and creates significantly adverse social effect due to violation of environmental protection laws and regulations in production and operation or some accidental factors. It mainly includes air pollution, radioactive pollutant, water pollution, noise and vibration pollution, pesticide and toxic chemical pollution.

（4）Obligation of obedience to regulation and supervision

Corporates are obliged to receive regulation and supervision from both the government and social public. In constructing projects, they should gain approval first before the commencement and perform the "Three Simultaneous System" amid the construction and disclose their environmental protection information to the public in a prompt way. For a construction project, it is forbidden to commence the construction in any of following three situations: (1) its environmental impact file is not submitted for approval as required; (2) its environmental impact assessment file is not re-submitted for a review as specified in Article 24, Environmental Impact Assessment Law; and (3) the re-submitted environmental impact assessment file is not approved by environmental protection administration or original approval division. According to Article 61 in the Environmental Protection Law, unauthorized commencement of a construction project that fails to pass environmental impact assessment review constitutes an offence against the law. Unlike other two obligations which are "soft" on enterprises, this is a compulsory one. As stipulated in Article 11 in the No. 13 Decree Methods on Environmental Protection Acceptance Check on Completed Construction Projects released by the State Environmental Protection Administration (SEPA) in 2001, construction projects should comply with "Three Simultaneous System" in classified environmental protection management of nationwide construction projects. When applying for as-built environmental protection acceptance check, the construction unit should submit following materials to competent environmental protection administration: as-built environmental protection acceptance application report and environmental protection acceptance monitoring report or survey report (for project with an environmental impact assessment report), as-built environmental protection acceptance check application form environmental protection acceptance monitoring report or survey report (for project with an environmental impact assessment form), or as-built environmental protection acceptance check registration card (for project with environmental impact registration form). The corporate environmental information disclosure obligation demands the corporate to disclose sewage discharge information and sewage alleviation facilities construction and operation to the social public and announce supervisory compliant channel so as to actively communicate with social public on any emerging issues.[18]

---

[18] Ji Luo, 'Development And Improvement Of The Cleaner Production System In China' (2001) 3 Chinese Journal of Population, Resources and Environment 11(in Chinese).

| Type | Details of obligation |
|---|---|
| Clean production obligation | • Enterprises should prioritize clean energy, process and equipment with higher resource utilization rate and less pollutant discharge, integrated waste utilization technology and pollutant detoxication process to reduce pollutant yield. |
| Obligation of emission reduction and lawful pollution discharge | • The enterprises and institutions should prevent and alleviate environmental pollution and eco-damage and undertake incurred liabilities.<br>• The enterprises, institutions and other operators involved in pollutant discharge should take measures to deal with environmental pollution and harm such as exhaust gas, wastewater, residue, medical waste, dust, foul gas, radioactive substance, noise, vibration, optical radiation, electromagnetic radiation resulting from their production, construction and other activities.<br>• It is strictly forbidden to illegally discharge pollutants through concealed pipe, seepage well, sewage pit, or pouring, falsify or forge monitoring data, or run pollution treatment facilities not as expected.<br>• The production, storage, transportation, sales, use and disposal of chemicals and radioactive substances should be consistent with relevant state rules and avoid contaminating the environment.<br>• The enterprises having a pollutant discharge permit should discharge pollutants as permitted; while those have not should not discharge pollutants at all.<br>• The enterprises, institutions and other operators containing pollutant discharge should pay effluent charge as required by the state. The effluent charge is exclusively used for environmental pollution prevention and treatment, which should not be retained, occupied or appropriated by any unit or individual. Units having paid environmental protection tax should be spared from effluent charge payment.<br>• The entities having polluted the environment and causing ecological damage should shoulder liability of infringement. |
| Environmental management obligation | • The enterprises and institutions containing pollutant discharge should establish environmental protection liability system to define the responsibilities of both their head and related staff.<br>• Enterprises containing significant pollutant discharge should install and use monitoring devices as required by the state and related regulations, make sure all the monitoring devices are operated normally, and keep original |

| | monitoring records.<br>• Enterprises and institutions should draft environmental contingency as required by related state laws and regulations and submit such contingency to competent environmental protection administration and other related bodies for filing. In case of an environmental emergency, the enterprise or institution involved should immediately take measures to deal with it and inform the units and residents that may be jeopardized, and submit a report to competent environmental protection administration and other related bodies. |
|---|---|
| Obligation of receiving regulation and supervision | • The construction project cannot be commenced unless an environmental impact assessment is completed.<br>•The pollution-preventing facilities required for a construction project must be designed, constructed and put into use synchronically with the major part. Such facilities should comply with the approved environmental impact assessment file and should not demolished or left in idle without being authorized.<br>•The units containing material pollutant discharge should faithfully disclose to the public the name of major pollutant, discharge pattern, discharge concentration and total volume and excessive discharge or not as well as the construction and operation of pollution prevention and treatment facilities. They should receive monitoring from the whole society. |

Table 1 Environmental obligations assumed by enterprises

**(II) Public jurisprudential structure of corporate environmental compliance: power-responsibility structure**

**1. Government environmental power**

The government environmental power refers to the power prescribed by laws and regulations for the government in terms of environmental protection. In accordance with Article 10, Environmental Protection Law, the government is entitled to implement environmental supervision on enterprises and perform environmental power. It is to supervise and manage the right of enterprises in utilizing the environmental resources so as to provide a pollution-and-damage free environment for the public. The power mainly consists of following aspects: (1) endow enterprises, institutions, social organizations and citizens to exploit all the natural resources owned by the state and discharge certain pollutants to the environment as an executor of state administrative functions; and (2) draft laws and regulations to protect and manage the environment and natural resources, punish the behaviors of jeopardizing the environment and unlawfully using environment and natural resources, maintain environmental

quality and functions, and protect citizens' environment right.[19]

Government environmental power corresponds to the corporate environmental obligation to some degree, thus the government is performing its environmental responsibility and regulating the relation between government environmental power and enterprise environmental power when exerting its environmental power. To be more specific, in the sense of independent utilization of environmental resources, environment utilization freedom and economic development freedom are unified. However, their correlation is not that simple. Modern economic development highlights the protection over freedom, whereas environmental protection stresses the importance of supervision. Such supervision aims to protect freedom instead of restricting it. In Jiangsu Province, local government once took "non-conventional measures" to shut down some enterprises so that the economic became stagnant and even showed the tendency of recession. With this incident as an example, we may find that excessive supervision may cause a "lose-lose" situation or even a situation with more than two losers. This makes it crystally clear that only the supervision in pursuit of freedom is fair. It is necessary to define the relationship between government power and market power as well as behavioral rules for both government and enterprises. The enterprises should be encouraged to pursue profit and supervised in operating behaviors and limited with adequate social responsibilities.

### 2. Corporate environmental responsibility

To achieve its ends, an enterprise must undertake certain environmental legal liabilities. Corporate environmental responsibility can be traced to the environmental obligation behind environmental right and environmental responsibility behind environmental power. Corporate environmental power indicates the power enjoyed by such organizations as enterprises to reasonably use environmental resources. It consists of two parts, namely the power of exploiting and utilizing natural resources and the power of emit or discharge pollutants within environmental tolerance (dumping or emission right). For the first part, it means natural resources owner and the entity having obtained the permit to develop, utilize, run and manage the state-owned natural resources are entitled to exploit such resources and get corresponding benefits. And for the second part, this power is also called dumping or emission power that is endowed by administrative authorities to the polluter to discharge pollutant or greenhouse gas to the environment within pollutant discharge (or control) limits prescribed by relevant laws and regulations.[20]

Based on the analysis of corporate environmental responsibility above, when there arises a conflict between shareholders' interest and social interest that may put public security at risk, it would be necessary for the state to exert its power to restrict the shareholders' interest. Some companies disregard ecological environmental protection and cause environmental pollution and ecological damage in maximizing their own interest. Out of the consideration for environmental interest of the society as a whole, the state may ask the corporate interest to give place to overall social interest and environmental protection need and apply some restrictions on the enterprises for protecting the environment and public interest. The Article 5 of the current Company Law offers a clear definition of corporate social responsibility. However, this is only a principle norm that serves for oath taking only without stipulating detailed contents of implementation. That's why companies could turn a blind eye to their social responsibility.[21] Since traditional social responsibility cannot exert effective supervision on the enterprises, drafting some hard pressing corporate responsibilities should be taken into consideration now. Corporate environmental responsibility consists of two parts: exploit and utilize natural resources and discharge pollutant within environmental tolerance. The responsibility of exploiting and utilizing natural resources is the power to develop and use such resources, which is accompanied with such responsibilities as protecting natural environment, maintaining ecological balance and exerting reasonable use, and managing and protecting the subject maters. By comparison, the responsibility of discharging pollutant within environmental tolerance is to discharge pollutant and emit greenhouse gas in accordance with the scope, methodology, routes, criteria, categories, concentration and quantity permitted by competent administrative agencies and perform environmental impact assessment, abide by the "Three Simultaneous System", declare and register pollution discharge, pay for effluent charge and receive onsite inspection.[22] The reason why the aforesaid corporate environment responsibilities should be called hard pressure rules is because the enterprises engaged in exploiting and utilizing environmental resources are compulsorily forced to balance environmental protection and economic development in their decision-making process. Obtaining economic interest should be accompanied with paying for corresponding environmental cost. In the meanwhile, when failing to comply with environmental protection laws and regulations, enterprises would be held liable for civil, administrative and criminal consequences. Nevertheless, an analysis of the private and public jurisprudences in corporate compliance reveals the environmental jurisprudence effective on corporate environmental compliance is not a simple private environmental law issue or a public environmental law issue. As a matter of fact, it is a complicated nested structure with public and private parts intertwined. The corporate environmental responsibility seems to transcend the category of obligation in private environmental law, because environmental law is a composite public-and-private law in its nature. Therefore, a novel environmental jurisprudence should be constructed to adapt to the environmental compliance situations. Below we're going to perform a specific demonstrative analysis of the environmental jurisprudence of this special structure.

---

[19] Zhongmei Lv, 'On the Comprehensive Decision Legal Mechanism of Ecological Civilization Construction'(2014) 3 China Legal Science(in Chinese).
[20] Mark Latham, Environmental Liabilities And The Federal Securities Laws: A Proposal For Improved Disclosure Of Climate Change-related Risks, 39 Envtl. L. 647.

### (III) Environmental jurisprudential structure of corporate environmental compliance: a dual layer nested structure

As suggested by the analysis above, enterprises are dealing with the government and social public in carrying out corporate compliance work. In the past, corporate environmental jurisprudence either incorporates enterprises into private law category for a separate analysis as per right-obligation structure in the civil law or includes it in the public law category for an analysis as per power-responsibility structure. It neglects the reality that enterprises are facing two situations in environmental compliance, ripping the real structure of corporate environmental jurisprudence. The fundamental reason for stagnation and crisis faced by environmental right theory is because scholars are trapped by the right-obligation paradigm in traditional private law. Their excessive reliance on traditional private law jurisprudence in analyzing the emerging environmental right brings in endless circular argumentation and self-contradiction. After realizing this problem, scholars make another mistake by overcorrecting it, directly denying the private right basis for environmental right and substituted the environmental right with environmental obligation. A bigger turmoil is brought to the field of environmental right theory. In fact, we should incorporate the environmental right theory into the category of public environmental law.[23]

---

[21] Lucia Ann Silecchia, Ounces of Prevention and Pounds of Cure: Developing Sound Policies for Environmental Compliance Programs, 7 Fordham Envtl. Law J. 583.

[22] David B. Spence, Paradox Lost: Logic, Morality, and the Foundations of Environmental Law in the 21st Century.

The application of the aforesaid two theories could only produce two consequences: campaign governance or laisser-faire governance. The legal interest of environmental law is neither private interest or public interest in the simple sense. It is actually a kind of public-and-private interest. What environmental jurisprudence is expected to illustrate should be an interacting pattern involving multi-layer and multi-behavior subjects. This interacting pattern results from interaction and communication among multiple acting subjects. Environmental law is never a one-way uniform response from either norm initiator or receiver but uniform interaction and communication between them. To be more specific, environmental law reflects social relationship developed during practical environment development, utilization, protection and management. Its subjectivity and comprehensiveness need to be tested in environmental practice. Summarization of the core idea of environmental law could only occur in environmental development, utilization, governance, protection and improvement but never in any imagination or books. Therefore, the authors construct a novel environmental jurisprudential structure faced by the enterprises in conducting environmental compliance work (see Fig. 2). As shown in the figure, enterprises have an access to some fundamental environmental rights which incur different environmental obligations. The obligations here correspond to the environmental rights enjoyed by the enterprises, including clean production, emission reduction, lawful pollution discharge, environmental management and obedience to regulation and supervision. Secondly, when enterprises are supervised by government environmental power and undertake corporate environmental responsibility, corporate environmental obligation becomes nested with corporate environmental responsibility. To put it simply, the corporate environmental obligation incurred by its environmental right should be within the scope of corporate environmental responsibility under government supervision. Finally, this structure embodying public-and-private interest of environmental law could explain the key point in corporate environmental compliance, namely enterprises could have an access to due environmental right on the basis of abiding by fundamental environmental obligation and government environmental supervision. This novel jurisprudential structure offers a convincing theoretical explanation for enterprises to carry out environmental compliance work and enables the enterprises to survive the fast-changing environmental regulatory environment. Based on this novel environmental jurisprudential structure, it is necessary to set up legal routes for current corporate environmental compliance so that the enterprises could be provided with a legal scheme to make sure about their compliance with the environmental law, coordination with government administrative regulation and communicating with the social public in environmental governance.

---

[23] Chunyu Zhu, 'A New Interpretation to Environmental Legal Relationships'(2018) 6 Journal of Zhengzhou University(Philosophy and Social Sciences Edition) 51(in Chinese).

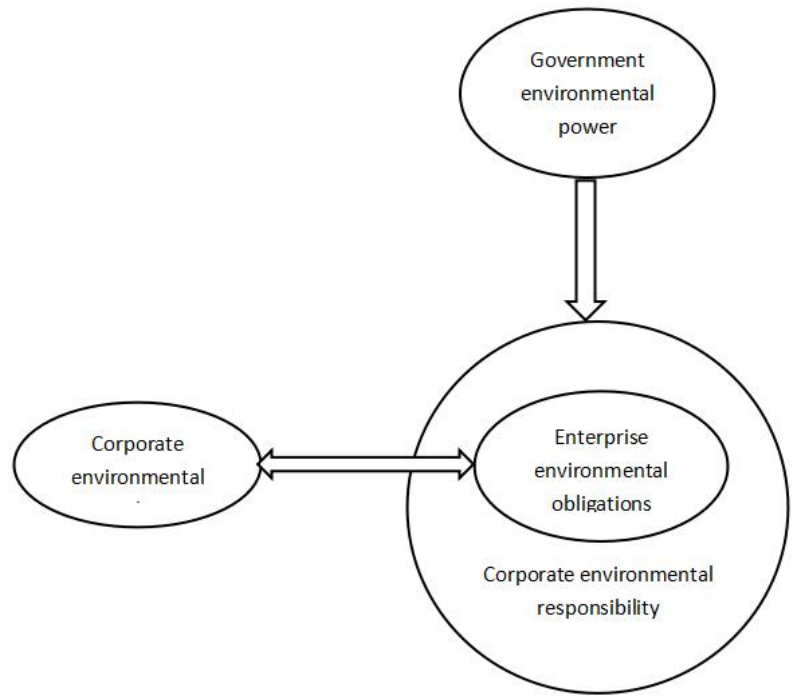

Fig. 1 A novel jurisprudential structure of corporate environmental compliance

## III. Routes of corporate environmental compliance: multi-party governance system

An effective execution of corporate environmental compliance is inseparable from a multi-party governance system that adapts well to Chinese practical conditions. In our past practice concerning environment governance, we depend overly on government supervision but neglect participation by the society and private entities so that the general public does not exert properly their environmental monitoring right. Now the high-pressure environmental supervision in the background of "carbon peaking and carbon neutrality" has further compressed the already small survival and development space left for the enterprises. Consequently, the enterprises have to seek survival and development in a highly adverse circumstance. In light of public-and-private environmental jurisprudence, the corporate environmental compliance could be divided into environmental compliance right and environmental compliance obligation. The former is an internal "soft-law" route along which corporate could exert self-monitoring in environmental self-discipline, whereas the latter is an external "hard-law" route along which entities other than enterprises supervise and monitor corporate environmental compliance. In the latter part, there is route of compliance with government supervision and that with environmental lawsuit on behalf of enterprises.

**(I) Incentive for environmental compliance right route: internal governance perspective**

Corporate environmental compliance is an important constituent part in accomplishing ecological legal system. For enterprises, how to establish an effective environmental legal incentive mechanism that can both ensure compliance with both external supervision and motivate the enterprises to spontaneously develop an internal compliance governance system is an important analytic thought in current study on environmental compliance legal routes exploration. In order to establish quite practical environmental compliance right incentive routes for enterprises, the authors have looked up to existing practices at home and beneficial experience at abroad, attempting to bridge corporate environmental compliance with corporate environmental interest by virtue of voluntary environmental information disclosure and environmental auditing so that enterprises have rules to follow in enjoying their environmental rights.

**1. Voluntary environmental information disclosure by enterprises**

Environmental information disclosure is an important route for accomplishing corporate environmental compliance. Due to asymmetry in environmental information, investors could not obtain such information promptly so that environmental market efficiency remains low. The inclusion of environmental information disclosure in the long existing and relatively developed securities disclosure mechanism could alleviate the asymmetry in environmental information and rise market efficiency.[24] In corporate environmental compliance practice among American enterprises, environmental obligation is deemed as consisting of three levels: (1) the state should not infringe on purpose the right of enterprises to carry out normal production and operation with environmental resources; (2) the state should protect the right enjoyed by the enterprises to carry out normal production and operation with environmental resources; and (3) the state should actively guarantee the execution of corporate environmental right by releasing favorable laws and policies. The reason why enterprises should be encouraged to establish their own information disclosure system is because of two factors. [25]Firstly, this could enhance enterprises' awareness of environmental compliance as well as its significance. And secondly, the corporate environmental compliance capacity could also be strengthened.[26] Against the increasingly complicated environmental laws and regulation, even those good-willed and diligent enterprises could benefit from environmental compliance work. Affected by market force, reduced waste treatment volume and better environmental protection reputation could enhance investors' interest and confidence. The enterprises are quite willing to comply with environmental laws and regulations without being sanctioned. The goal for enterprises to establish voluntary environmental information disclosure system is quite clear, namely to motivate enterprises to voluntarily disclose their environmental information by virtue of market incentive measures.[27]

---

[24] John W. Bagby, Paula C. Murray, Eric T. Andrews, SHOW GREEN WAS MY BALANCE SHEET?: CORPORATE LIABILITY AND ENVIRONMENTAL DISCLOSURE, 14 Va. Envtl. L.J. 225.
[25] Peter J. Fontaine, EPA's Multimedia Enforcement Strategy: The Struggle to Close the Environmental Compliance Circle,18 Colum. J. Envtl. L. 31.

Based on the analysis above, we suggest following routes for voluntary corporate environmental information disclosure: (1) draft and release specific corporate environmental protection policies and annual environmental protection objectives and efficacy; (2) announce annual total energy consumed; (3) disclose environmental protection investment and technical development; (4) disclose the types, quantities, concentrations and whereabouts of the pollutants discharged; (5) disclose data concerning construction and operation of environmental protection facilities; (6) disclose the treatment and disposal of wastes resulting from production activities as well as waste recycle and integrated use; (7) disclose the voluntary agreement about environmental amelioration behaviors signed with environmental protection administrations; (8) disclose rewards from environmental protection administrations; (9) disclose other information as possible, including specific items concerning environment in the financial statements, detailed environmental cost accounting, remarks concerning environmental information sources. All those efforts should be made to eliminate the significant gap in environmental information between enterprises and investors.

**2. Environmental accounting compliance by enterprises**

The key of environmental accounting compliance lies in collecting related environmental data, information and records and analyzing their compliance with relevant laws, regulations and supervision technologies so as to figure out the real environmental health conditions of the enterprises. However, data collection cannot be limited to corporate records and files; or there is risk of being suspected as having committed forgery. Files should be registered at environmental protection administrations and both state and local environmental protection laws and regulations should be complied with. Besides, no punishment should be based on the voluntary environmental accounting compliance findings from the enterprises.[28] Construction of corporate environmental accounting compliance routes needs common efforts from the government, enterprises and social public. Corporate accounting compliance is directly accessible for the government and social public. The social public is the direct victims of environmental pollution; thus, they have strong motive to monitor environmental duty implementation by enterprises. With "public appeal" mechanism that could make sure about public participation in environmental accounting compliance being established, the general public could determine whether and how the enterprises comply with environmental protecting policies and systems through the publicly disclosed environmental compliance accounting report. Apart from public monitoring of corporate environmental liability implementation through public participation mechanism, the government should also properly instruct the enterprises to join environmental compliance accounting and offer professionals for helping those enterprises in planning and adjusting environmental accounting work. These measures could lower corporate environmental risk and help enterprises in really performing their environmental responsibilities.[29]

---

[26] David Waskow, Chris Jochnick & Juge Gregg, Progressive Approaches to International Environmental Compliance, 15 FORDHAM ENVTL. L. REV. 496 (2004).
[27] Timothy Riley and Cai Huiyan, UNMASKING CHINESE BUSINESS ENTERPRISES: USING INFORMATION DISCLOSURE LAWS TO ENHANCE PUBLIC PARTICIPATION IN CORPORATE ENVIRONMENTAL DECISION MAKING, 33 Harv. Envtl. L. Rev. 177.
[28] Somendu B. Majumdar, Voluntary Environmental Compliance Auditing: A Primer, 7 Fordham Envtl. Law J. 817.

### (II) Environmental compliance control obligation: external supervision perspective

Enterprises usually moves passively and slowly in front of climate changes, environment contamination and ecological damage. They are facing a string of environmental risks, including government regulator risk and lawsuit risk, so that some of them have to pay tremendous costs such as civil compensation, administrative fine or even criminal sanction. Corporate environmental compliance stands out as a countermeasure in dealing with environmental risks. Enterprises should not only exert self-discipline from environmental compliance rights but also find routes of environmental compliance from their obligations. Both substantial matters and procedural matters involved should be figured out so that the enterprises could develop active environmental risk response plan and better accomplish corporate environmental compliance.[30]

**1. Routes for compliance with government regulation**

How to cope with government regulation is an important part in corporate environmental compliance. For enterprises, corporate environmental compliance largely relies on finding out which environmental laws and regulations and when those are applicable to them as well as what relations those applicable clauses have with each other and what systems they are from.[31] To comply with government regulation, enterprises should make efforts from two levels. Firstly, they should make sure internal environmental compliance is consistent with all the environmental regulatory laws and regulations of the government and cooperate with government regulation. Secondly, it is necessary to strengthen their environmental socio-governance (ESG) system so that they could better deal with external supervision.

First of all, corporate environmental compliance should be consistent with environmental regulatory laws and regulations. As shown in Table 1, enterprises should pay attention to environmental resource carrying capacity compliance and environmental resource permit compliance. Environmental resources bearing capacity compliance means enterprises should stay within the self-maintenance and self-regulation capacity of environmental system during the development. Attention should be paid to the supply and tolerance capacities of resources and environmental subsystem. Economic development intension should be reasonable. This could be conducted from three aspects, namely corporate environmental impact assessment, corporate water assessment and corporate energy demand assessment. As to the environmental resources permit compliance, it means enterprises must obtain related permits before assigning environmental resources available, including sewage permit, water use permit and mining permit.[32]

---

[29] Rena I. Steinzor, Reinventing Environmental Regulation: The Dangerous Journey From Command To Self-Control, 22 Harv. Envtl. L. Rev. 103
[30] Mark Latham, ENVIRONMENTAL LIABILITIES AND THE FEDERAL SECURITIES LAWS: A PROPOSAL FOR IMPROVED DISCLOSURE OF CLIMATE CHANGE-RELATED RISKS, 39 Envtl. L. 647.
[31] Richard F. Charfield-Taylor, Environmental Compliance: Negotiating the Regulatory Maze, 3 MO. ENVTL. L. & POL'y REV. 3 (1995).

| Route | Specific type | Legal ground |
|---|---|---|
| Environmental resources bearing capacity compliance route | Environmental impact assessment | Article 2 and 16, Environmental Impact Assessment Ordinance |
| | Water assessment | Article 51 and 53, Water Act |
| | Energy demand assessment | Article 24, 25, 26 and 27, Energy Conservation Law |
| Environmental resources permit compliance route | Pollutants discharge permit | Article 45, Environmental Protection Law; Article 21, Water Pollution Control Act; Article 19, Air Pollution Prevention and Control Act |
| | Water use permit | Article 2, Administrative Regulations on Water Permit and Water Resources Fee Levy; Article 5 and 7, Regulations on Water Permit |
| | Mining permit | Article 3, 29, 30 and 32, Mineral Resources Law |

Table 2 Routes for compliance with government environmental regulation by enterprises

Furthermore, ESG should be improved to help the enterprises in carrying out corporate environmental compliance work in an orderly way. ESG construction indicates institutionalization of corporate environmental social responsibility. There are two routes, namely "hard-law" and "soft-law". The so-called "hard-law" route is to legalize corresponding responsibilities, which directs at moral bottom line. Coercive means would be taken to bound the enterprises and prevent them from damaging the environmental interest of the public and interested parties. By comparison, the "soft-law" route corresponds to "socialization". It is more about the active ones in corporate environmental social responsibilities. Multiple parties would be invited to guide and encourage the enterprises to promote the environmental interests of both the public and interested parties. To be more specific, enterprises should enhance ESG in coping with government environmental regulation risk and gradually improve the environment and social environmental value discovery and environmental reputation incentive mechanism so as to cooperate with government's environmental regulation on them.[33]

---
[32] Eleventh Annual Symposium: Environmental Protection In The Developing World: A Look At The Responsibility Of State And Non-State Actors, 15 Fordham Envtl. Law Rev. 403.

### 2. Compliance routes for enterprises in dealing with environmental lawsuits

Social public and their representing institutions or organizations constitute important entities that may file an environmental lawsuit against enterprises. It is really important for enterprises to carry out environmental compliance work in dealing with environmental lawsuit risk. The environmental lawsuit risk hanging over enterprises may come from the procuratorate (criminal lawsuit) and social public and environmental protecting organizations (civil lawsuit). Enterprises should draft environmental compliance plans according to the classification of lawsuits, namely criminal compliance and civil compliance.

（1）Compliance routes for dealing with criminal lawsuit

Now more and more enterprises are seeking the approaches that could help them avoid environmental criminal lawsuit. A possible solution is to carry out voluntary information disclosure and environmental accounting.[34] As a lawful plan that could strengthen crime prevention and improve enterprise situation, corporate environmental criminal compliance is a "duty of care" voluntarily established by enterprises and their internal supervisory departments instead of a "statutory duty". Future development of corporate environmental criminal compliance should be based on domestic existing theoretical studies on institutional crimes. Enterprises should be guided to prevent and control environmental crimes by referring to related environmental criminal policies so that compliant enterprises could be freed and socio-economic development could be promoted. Specifically speaking, prerequisites for enterprises to be spared from prosecution with criminal compliance include: (1) keep promises about corporate environmental compliance; (2) active cooperate with related parties and report about corporate environmental conditions; (3) be determined as subjectively negligent in case of unlawful act; and (4) immediately take measures to help with victims.[35] Thus, the specific routes for enterprises to conduct environmental compliance work are listed as follows: in the first place, enterprises should sign an environmental compliance promise with the procuratorate in charge of pressing criminal lawsuits and actively fulfill it; then, enterprises should proactively work together with the procuratorate and environmental protecting organizations in voluntarily disclosing environmental information and environmental accounting compliance work; next, they should make every possible effort to collate, file and keep environmental compliance materials for testifying subjective negligence when accused of committing any unlawful act; and at last, enterprises should take active measures to help those victims of environmental damage and negotiate about conflict resolution.[36]

---

[33] Somendu B. Majumdar, Voluntary Environmental Compliance Auditing: A Primer, 7 Fordham Envtl. Law J. 817.
[34] Nancy K. Kubasek, M. Neil Browne & Carrie Williamson, Role of Criminal Enforcement in Attaining Environmental Compliance in the United States and Abroad, 7 U. BALT. J.
[35] Donald A. Carr and William L. Thomas, Devising a Compliance Strategy Under the ISO 14000 International Environmental Management Standards, 15 Pace Envtl. L. Rev. 85.

（2）Compliance routes for dealing with civil lawsuit

Environmental civil compliance routes are strategies that should be taken by enterprises in dealing with lawsuits filed by social public or environmental protection organizations. The fundamental thought is to negotiate with possibly interested parties and reach an agreement on compensation plan before development and utilization of corresponding environmental resources.[37] This is also called compensation before development. The order cannot be reversed anyway. The routes here could be divided into public lawsuit compliance route and private lawsuit compliance route. The division is based on the difference in litigants faced by the enterprises. We could start with public-interest civil lawsuit compliance. In such case, enterprises are dealing with such litigants as environmental protection organizations or the procuratorate. It accounts for the necessity of drafting two different schemes. When a public-interest environmental lawsuit is filed by an environmental protection organization, negotiation should be held with the organization to reach agreement on compensation matters and a reconciliation out of litigation. This could well maintain the corporate reputation. When the public-interest environmental lawsuit is pressed by a procuratorate, the enterprise involved should cooperate with the procuratorate's investigation, actively submit prepared environmental compliance materials and reports, strive for a reconciliation out of litigation, and perform the reconciliation agreement as promised. [38] Besides, there is also private-interest civil environmental lawsuit. In this situation, enterprises are facing the social public with damaged environmental interest. Thus, the enterprises could take plea against the civil lawsuit so as to alleviate their responsibilities. In addition, since mass disturbance may happen after social public suffers a damage in environmental rights and interests, enterprises involved should provide preliminary compensation, propaganda and comfort work to eliminate the other party's contradicting sentiment and conflicts.[39]

As shown in Fig. 3, future corporate environmental compliance should form a multi-party governance pattern that is guided by the government, participated by corporates and supervised by the society. In environmental compliance work, enterprises should give a full play to their initiative. In addition to actively performing environmental protection obligation, they should voluntarily set up corporate environmental governance rules, explore and improve voluntary environmental compliance system, and work together with government's compulsory environmental information disclosure and social public's environmental monitoring system. All those measures aim to form a multi-party governance pattern that could guarantee orderly production and operation for enterprises on the premise of abiding by environmental protection laws and regulations.

---

[36] Nancy K. Kubasek, M. Neil Browne & Carrie Williamson, Role of Criminal Enforcement in Attaining Environmental Compliance in the United States and Abroad, 7 U. BALT. J. ENVTL. L. 122 (2000).
[37] Stephanie Stimpson, Strategies For Risk Management And Corporate Social Responsibility For Oil And Gas Companies In Emerging Markets, 53 Alberta L. Rev. 259.
[38] Andrew I. Davis, Judicial Review of Environmental Compliance Orders, 24 ENVTL. L. 189(1994).
[39] John W. Bagby, Show Green Was My Balanc Sheet?: Corporate Liability And Environmental DisclosureI, 14 Va. Envtl. L.J. 225.

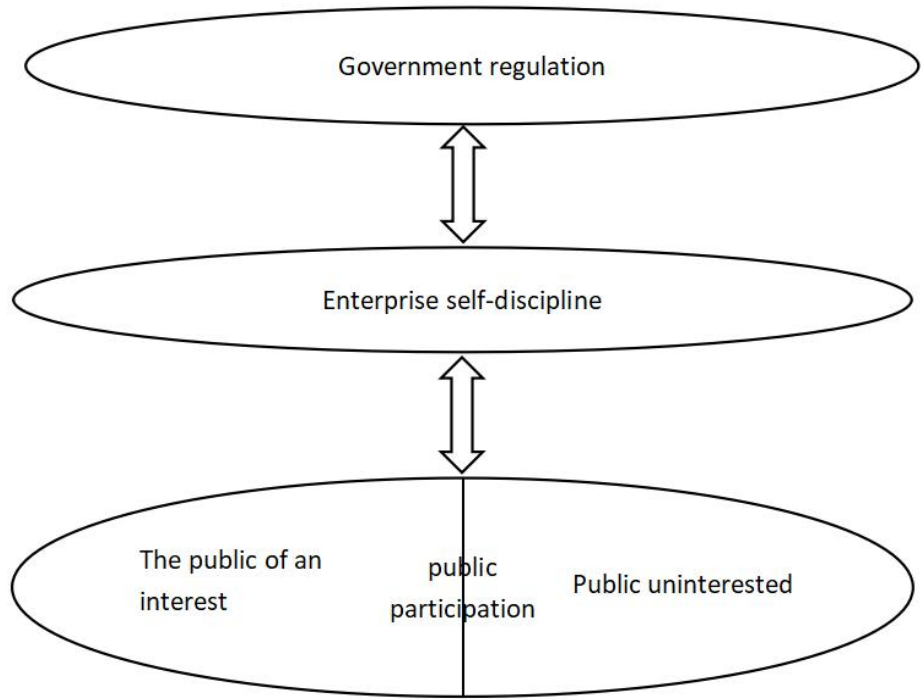

Fig. 2 Multi-party governance system involving duly functioning government, enterprises and social public

# Conclusion

It is of due significance to improve corporate environmental compliance mechanism and develop an environmental governance system built, governed and shared jointly among the government, enterprises and the social public for modernizing eco-environment governance system and governance capacity. Against the background of "carbon peaking and carbon neutrality" goals, enterprises should deal with high-pressure environmental regulation from the government and solve environmental disputes and conflicts with the social public. Two types of risks from government regulation and environmental lawsuit, respectively, are hanging over the enterprises. Corporate environmental compliance should be based on a novel environmental jurisprudence of dual layer nested structure. Such structure could illustrate both due rights and obligations of enterprises in their production and operation activities and environmental power and responsibilities of the government in supervising the enterprises. Corporate obligation is nested within corporate responsibility system. In light of such environmental jurisprudence, legal accomplishment routes for constructing corporate environmental compliance should be divided into two kinds, namely right-related route and obligation-related route. The environmental compliance right-related routes correspond to internal governance routes, according to which enterprises are entitled to environmental compliance routes extended from their own environmental rights. By contrast, the environmental compliance obligation-related routes are about external regulatory and supervision routes, including those for them to cope with government supervision and environmental lawsuits. The ultimate objective is to form a multi-party governance system featuring government supervision, corporate self-discipline and social public participation. This study may shed some light on the theoretical studies and implementation schemes concerning corporate environmental compliance.

# References

[1]David Waskow, Chris Jochnick & Juge Gregg, Progressive Approaches to International Environmental Compliance, 15 FORDHAM ENVTL. L. REV. 496 (2004).

[2]Richard F. Charfield-Taylor, Environmental Compliance: Negotiating the Regulatory Maze, 3 MO. ENVTL. L. & POL'y REV. 3 (1995).

[3]Wanyi Zhao, Peng Wang, 'On the Legal Realization Route of Synthetical and Harmonized Regulating Corporates Compliance Behaviors in China' (2021) 7 Hebei Law Science 39(in Chinese).

[4]Ruihua Chen, Basic Theory of Enterprise Compliance(2nd edn, Law Press 2021)103-65(in Chinese).

[5]Timothy Riley and Cai Huiyan, Unmasking Chinese Business Enterprises: Using Information Disclosure Laws To Enhance Public Participation In Corporate Environment Decision Making, 33 Harv. Envtl. L. Rev. 177.

[6]Yang Chen, 'On the logic of check and balance in China's environmental multi-governance',(2020) 6 Chinese Journal of Population, Resources and Environment 30(in Chinese).

[7]Bowen, H. R . Social Responsibilities of the Business Man [J]. Harper Brothers, 1953(6).

[8]Binglin Tan, 'On the Government's Management-Based Regulation of Enterprises' (2019) 6 Journal of Jurists(in Chinese).

[9]Donald A. Carr, William L. Thomas, Devising a Compliance Strategy Under the ISO 14000 International Environmental Management Standards, 15 Pace Envtl. L. Rev. 85.

[10]Qian Zhu, Jingjing Yu, 'Research on the approach to the Connection of Environmental Civil Public and Private Interest Litigation from the Perspective of Civil Code' (2021) 5 Journal of China University of Geosciences (social sciences edition) 21(in Chinese).

[11]Yixiang Xu, 'The binary structure of public participation rights: Focusing on the legal norms of public participation in the environmental administration' (2018) 2 Journal of Central South University (Social Sciences Edition) 24(in Chinese).

[12]Qian Zhu, Jingjing Yu, 'Research on the approach to the Connection of Environmental Civil Public and Private Interest Litigation from the Perspective of Civil Code' (2021) 5 Journal of China University of Geosciences (social sciences edition) 21(in Chinese).

[13]Wenxian Zhang, Research on the Category of Legal Philosophy(1st edn, China University of Political Science and Law Press 2001 )399-281(in Chinese).

[14]Rena I. Steinzor, REINVENTING ENVIRONMENTAL REGULATION: THE DANGEROUS JOURNEY FROM COMMAND TO SELF-CONTROL, 22 Harv. Envtl. L. Rev. 103.

[15]Somendu B. Majumdar, Voluntary Environmental Compliance Auditing: A Primer, 7 Fordham Envtl. Law J. 817.

[16]Donald A. Carr and William L. Thomas, Devising a Compliance Strategy Under the ISO 14000 International Environmental Management Standards, 15 Pace Envtl. L. Rev. 85.

[17]Shouqiu Cai,'Exploration of Environmental Rights'(1982) 3 Chinese Social Sciences(in Chinese).

[18]Ji Luo, 'Development And Improvement Of The Cleaner Production System In China' (2001) 3 Chinese Journal of Population, Resources and Environment 11(in Chinese).

[19]Zhongmei Lv, 'On the Comprehensive Decision Legal Mechanism of Ecological Civilization Construction' (2014) 3 China Legal Science(in Chinese).

[20]Mark Latham, Environmental Liabilities And The Federal Securities Laws: A Proposal For Improved Disclosure Of Climate Change-related Risks, 39 Envtl. L. 647.

[21]Lucia Ann Silecchia, Ounces of Prevention and Pounds of Cure: Developing Sound Policies for Environmental Compliance Programs, 7 Fordham Envtl. Law J. 583.

[22]David B. Spence, Paradox Lost: Logic, Morality, and the Foundations of Environmental Law in the 21st Century.

[23]Chunyu Zhu, 'A New Interpretation to Environmental Legal Relationships' (2018) 6 Journal of Zhengzhou University(Philosophy and Social Sciences Edition) 51(in Chinese).

[24]John W. Bagby, Paula C. Murray, Eric T. Andrews, SHOW GREEN WAS MY BALANCE SHEET?: CORPORATE LIABILITY AND ENVIRONMENTAL DISCLOSURE, 14 Va. Envtl. L.J. 225.

[25]Peter J. Fontaine, EPA's Multimedia Enforcement Strategy: The Struggle to Close the Environmental Compliance Circle,18 Colum. J. Envtl. L. 31.

[26]David Waskow, Chris Jochnick & Juge Gregg, Progressive Approaches to International Environmental Compliance, 15 FORDHAM ENVTL. L. REV. 496 (2004).

[27]Timothy Riley and Cai Huiyan, UNMASKING CHINESE BUSINESS ENTERPRISES:

USING INFORMATION DISCLOSURE LAWS TO ENHANCE PUBLIC PARTICIPATION IN CORPORATE ENVIRONMENTAL DECISION MAKING, 33 Harv. Envtl. L. Rev. 177.

[28]Somendu B. Majumdar, Voluntary Environmental Compliance Auditing: A Primer, 7 Fordham Envtl. Law J. 817.

[29]Rena I. Steinzor, Reinventing Environmental Regulation: The Dangerous Journey From Command To Self-Control, 22 Harv. Envtl. L. Rev. 103

[30]Mark Latham, ENVIRONMENTAL LIABILITIES AND THE FEDERAL SECURITIES LAWS: A PROPOSAL FOR IMPROVED DISCLOSURE OF CLIMATE CHANGE-RELATED RISKS, 39 Envtl. L. 647.

[31]Richard F. Charfield-Taylor, Environmental Compliance: Negotiating the Regulatory Maze, 3 MO. ENVTL. L. & POL'y REV. 3 (1995).

[32]Eleventh Annual Symposium: Environmental Protection In The Developing World: A Look At The Responsibility Of State And Non-State Actors, 15 Fordham Envtl. Law Rev. 403.

[33]Somendu B. Majumdar, Voluntary Environmental Compliance Auditing: A Primer, 7 Fordham Envtl. Law J. 817.

[34]Nancy K. Kubasek, M. Neil Browne & Carrie Williamson, Role of Criminal Enforcement in Attaining Environmental Compliance in the United States and Abroad, 7 U. BALT. J.

[35]Donald A. Carr and William L. Thomas, Devising a Compliance Strategy Under the ISO 14000 International Environmental Management Standards, 15 Pace Envtl. L. Rev. 85.

[36]Nancy K. Kubasek, M. Neil Browne & Carrie Williamson, Role of Criminal Enforcement in Attaining Environmental Compliance in the United States and Abroad, 7 U. BALT. J. ENVTL. L. 122 (2000).

[37]Stephanie Stimpson, Strategies For Risk Management And Corporate Social Responsibility For Oil And Gas Companies In Emerging Markets, 53 Alberta L. Rev. 259.

[38]Andrew I. Davis, Judicial Review of Environmental Compliance Orders, 24 ENVTL. L. 189(1994).

[39]John W. Bagby, Show Green Was My Balanc Sheet?: Corporate Liability And Environmental DisclosureI, 14 Va. Envtl. L.J. 225.